\documentclass[11pt]{article}

\usepackage[T1]{fontenc}
\usepackage[utf8]{inputenc}
\usepackage{lmodern}
\usepackage[letterpaper,margin=1in]{geometry}
\usepackage{microtype}
\usepackage{booktabs}
\usepackage{array}
\usepackage{tabularx}
\usepackage{longtable}
\usepackage{amsmath}
\usepackage{enumitem}
\usepackage{xcolor}
\usepackage{graphicx}
\usepackage{float}
\usepackage{tikz}
\usetikzlibrary{arrows.meta,positioning,fit,backgrounds,calc}
\usepackage{caption}
\usepackage{xurl}
\usepackage[hidelinks]{hyperref}
\usepackage{titlesec}
\usepackage{needspace}
\usepackage{claims}

\definecolor{ink}{HTML}{172033}
\definecolor{muted}{HTML}{5D687B}
\definecolor{accent}{HTML}{1261A0}
\definecolor{danger}{HTML}{B93A32}
\definecolor{safe}{HTML}{237A57}
\definecolor{panel}{HTML}{F3F6F9}

\hypersetup{
  pdftitle={Agent Name Collision Attacks in Multi-Agent Systems},
  pdfauthor={Adithyan Arun Kumar},
  pdfsubject={Security analysis of name-derived routing identity in multi-agent systems},
  pdfkeywords={A2A, multi-agent systems, identity, routing, name collision, agent security}
}
\renewcommand{\arraystretch}{1.20}
\newcolumntype{L}[1]{>{\raggedright\arraybackslash}p{#1}}
\newcolumntype{Y}{>{\raggedright\arraybackslash}X}
\titleformat{\section}{\normalfont\Large\bfseries}{\thesection}{0.75em}{}
\titleformat{\subsection}{\normalfont\large\bfseries}{\thesubsection}{0.75em}{}
\newcommand{\code}[1]{\texttt{#1}}

\newcommand{\QualifiedImplementations}{7}
\newcommand{\MaintainerOrganizations}{6}
\newcommand{\ClientDispatches}{6}
\newcommand{\BrokerCollisions}{1}
\newcommand{\DirectCredentialTransfers}{0}
\newcommand{\ConditionalDelegatedCases}{1}
\newcommand{\ModelMediatedChains}{2}
\newcommand{\OfficialSamplePaths}{6}
\newcommand{\PropagationRepositories}{83}
\newcommand{\NegativeControls}{7}

\title{Agent Name Collision Attacks in Multi-Agent Systems}
\author{Adithyan Arun Kumar\\
Independent Security Researcher\\
\href{mailto:adioffsec@gmail.com}{\texttt{adioffsec@gmail.com}}}
\date{September 2026}

\begin{document}
\maketitle

\begin{abstract}
Multi-agent hosts turn remote Agent Cards into local agents, tools, workflow targets, and broker routes. A2A defines the card's \code{name} as human-readable metadata, not as a stable identity, and specifies no collision semantics. The security failure begins when a host nevertheless uses that remote name as a local routing identifier.
We traced registration through dispatch and ran isolated regression tests at seven pinned open-source revisions. Six client-style integrations selected an attacker-controlled peer's client or loopback endpoint for a request addressed to a trusted peer's name. A seventh, brokered implementation collapsed both peers onto one name-derived route; queue and access-control state determine whether the result is interception or denial. The common result is wrong-peer dispatch, not universal privilege inheritance. Synthetic credential and tool tests found no A-specific credential transfer in the tested client bindings and no direct transfer of A-owned tools. The broker path forwards a caller-configuration object; delegated identity or tokens reach B only if present and B can consume the route. Two other paths expose a later, model-mediated decision rather than direct execution authority.
The necessary conditions assign different responsibilities to the protocol, implementations, and deployments. Hosts should route by an origin-bound stable identity, keep names presentational, and reject ambiguous aliases. The evidence establishes a recurring implementation vulnerability class, not a universal A2A protocol exploit or a count of vulnerable deployments.
\end{abstract}

\section{Introduction}

Agent discovery converts a remote description into a local object that can receive tasks. The host starts with an endpoint, registry record, authenticated subject, or broker principal and then retrieves an Agent Card. A secure design keeps that admitted identity bound to the transport used at dispatch.
\claim{C002}The failure begins when a host takes a peer-controlled display name, uses it as a local agent, tool, workflow, registry, or broker selector, accepts a collision, and silently chooses one peer by order or shared route. This resolves a name to a resource outside the intended control sphere, consistent with CWE-706\cite{cwe-706}.

The instruction ``send this task to \code{payments}'' can appear to preserve the operator's choice even after the resolver has rebound \code{payments} to another principal. The attacker does not need to forge A's transport, signing key, or credentials. It needs control of an admitted peer B's display name and a resolver that lets B win the collision.

\claim{C003}We evaluated \QualifiedImplementations{} open-source implementations maintained across \MaintainerOrganizations{} organizations at exact revisions. The corpus spans agent-tree lookup, generated function tools, client maps, workflow registries, and broker topics. It is purposive rather than statistically representative. The result is breadth across independently implemented sinks, not an estimate of deployment prevalence. The affected source paths are pinned in the repositories for Google ADK Python and TypeScript, UiPath LangChain, BeeAI Framework, Solace Agent Mesh, Mozilla Any-Agent, and AutoDev\cite{adk-python-b018062,adk-js-ac84bb4,uipath-langchain-5331f6e,beeai-0e7e3e5,solace-mesh-f71e744,any-agent-58938fe,autodev-23777fe}.
\claim{C004}In controlled tests, \ClientDispatches{} client-style integrations selected B's client or loopback endpoint for an A-selected request. The brokered target produced \BrokerCollisions{} name-derived route collision; its final delivery outcome depends on broker configuration. These results establish wrong-peer dispatch, not universal credential theft or code execution.

This work contributes:

\begin{enumerate}
  \item a precise identity invariant and a conjunctive threat model that distinguish an exploitable collision from a harmless duplicate;
  \item registration-to-dispatch traces and controlled tests across seven implementation paths, including a separate authority-transfer analysis;
  \item an evidence-calibrated division between protocol semantics, implementation behavior, and deployment admission; and
  \item a fail-closed design in which routing uses a stable identity bound to enrollment origin while names remain presentation metadata.
\end{enumerate}

The evidence supports a recurring implementation vulnerability enabled by an identity-semantics gap in A2A. It does not show a break in wire framing, automatic enrollment, vulnerability in every A2A SDK, or seven remotely exploitable production deployments.

\Needspace{14\baselineskip}
\section{System and Threat Model}

\subsection{Identity layers}

Each admitted peer has a stable identity bound to its admitted origin or transport. A remotely supplied card name is presentation metadata; it must not replace or modify that binding when the host resolves a request.

\claim{C002}A collision becomes a vulnerability when the host instead uses the card name as an authority-bearing selector, permits two distinct peers to share it, and then uses the ambiguous value to select a transport or execution object. For a trusted A and attacker-controlled B with the same name, the selected object can depend on registration order, list order, normalization, or broker binding state rather than the identity selected at admission.

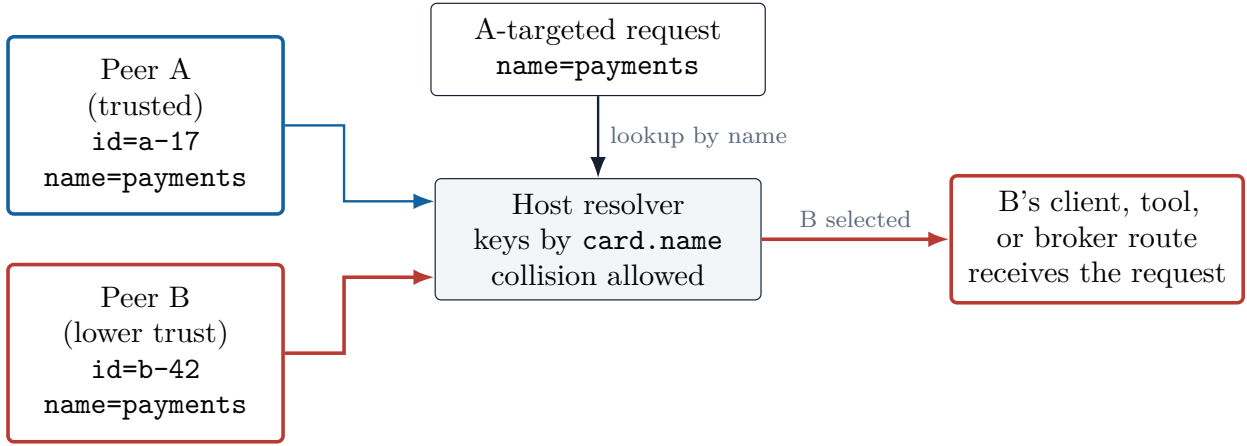
\begin{figure}[H]
  \centering
  \resizebox{\textwidth}{!}{\begin{tikzpicture}[
  x=1cm,
  y=1cm,
  font=\small,
  box/.style={draw=ink,rounded corners=2pt,align=center,minimum height=10mm,inner sep=5pt,fill=white},
  trusted/.style={box,draw=accent,very thick},
  attacker/.style={box,draw=danger,very thick},
  host/.style={box,draw=ink,fill=panel,text width=35mm},
  arr/.style={-{Latex[length=2.5mm]},thick,draw=ink},
  trustedarr/.style={-{Latex[length=2.5mm]},thick,draw=accent},
  bad/.style={-{Latex[length=2.5mm]},very thick,draw=danger},
  note/.style={font=\scriptsize,color=muted,align=center}
]
  \node[trusted,text width=29mm,minimum height=21mm] (a) at (0,1.35) {Peer A\\(trusted)\\\code{id=a-17}\\\code{name=payments}};
  \node[attacker,text width=29mm,minimum height=21mm] (b) at (0,-1.35) {Peer B\\(lower trust)\\\code{id=b-42}\\\code{name=payments}};
  \node[host] (registry) at (5.35,0) {Host resolver\\keys by \code{card.name}\\collision allowed};
  \node[box,text width=36mm] (request) at (5.35,2.25) {A-targeted request\\\code{name=payments}};
  \node[attacker,text width=31mm] (sink) at (11.25,0) {B's client, tool,\\or broker route\\receives the request};

  % The two bent paths approach the resolver horizontally so their arrowheads
  % terminate squarely on its west border rather than pointing up or down.
  \draw[trustedarr] (a.east) -- ++(7mm,0) |- ([yshift=4.5mm]registry.west);
  \draw[bad] (b.east) -- ++(7mm,0) |- ([yshift=-4.5mm]registry.west);
  \draw[arr] (request.south) -- node[right,note]{lookup by name} (registry.north);
  \draw[bad] (registry.east) -- node[above,note]{B selected} (sink.west);
\end{tikzpicture}}
  \caption{A name collision breaks the binding between the admitted peer and the selected route. The resolver accepts both cards under one display name and sends A's request to B.}
  \label{fig:attack-flow}
\end{figure}

The selector can be a dictionary key, a tree lookup, a generated function-tool name, a workflow node, a request topic, or a queue identity. The data structure is incidental. The invariant fails at the point where presentation metadata determines which principal receives a task.

\subsection{Attacker capabilities and necessary conditions}

\claim{C005}A confidentiality or integrity attack requires all of the following:

\begin{enumerate}
  \item \textbf{Shared routing domain.} A and B are both admitted to the same host, registry, tool set, workflow, or broker namespace. The name collision does not itself enroll B.
  \item \textbf{Name control.} B can publish or refresh its own card, or an attacker has compromised an admitted B and can change its card name.
  \item \textbf{Collision precedence.} B is placed first in a first-match resolver, registered last in a replacement map, normalized into A's tool identifier, or able to bind or consume the broker route.
  \item \textbf{Name-derived use.} A caller, workflow, or model later selects the ambiguous local name. A client that sends directly to a fixed endpoint has no collision surface.
  \item \textbf{Trust differential.} A may receive data or exercise authority that B may not. If A and B are fully interchangeable for every routed task, the impact can shrink to correctness or availability.
\end{enumerate}

The attacker is inside the agent admission boundary but remains outside A and the host. This model covers registries, marketplaces, partner integrations, multi-account orchestration, and brokered meshes. It does not cover an unauthenticated Internet attacker.

\subsection{Security outcomes}

We separate four outcomes that are often collapsed in vulnerability descriptions:

\begin{description}[style=nextline]
  \item[Message diversion] B receives task or caller context intended for A and may respond on the selected route.
  \item[Credential transfer] A-specific secret material appears at B's request boundary.
  \item[Authority transfer] B receives a delegated identity or token, an A-owned tool object, or an equivalent direct capability.
  \item[Secondary execution] B's response influences a later model or policy decision that may invoke a host tool.
\end{description}

Only the first follows directly from client-side wrong-peer dispatch. The remaining outcomes depend on what the host attaches after routing and what decisions occur after B responds.

\section{Methodology}

\subsection{Inclusion criteria}

We retained a public-source implementation only when all six conditions held:

\begin{enumerate}
  \item normal framework or product code accepts multiple remote agents or cards;
  \item the peers remain distinguishable by endpoint, resource ID, registry record, or equivalent input;
  \item a remote card name becomes an authority-bearing local selector;
  \item exact or normalization-equivalent duplicates do not fail closed;
  \item a later production path resolves through that selector; and
  \item a focused test or complete source trace reaches the selected client, endpoint, tool, topic, or queue.
\end{enumerate}

A dictionary assignment or code-search hit was insufficient. Copied samples were not counted as independent implementations. \claim{C003}These criteria retained \QualifiedImplementations{} implementations across \MaintainerOrganizations{} maintainer organizations. Table~\ref{tab:targets} gives the exact revisions and evidence types.

\subsection{Controlled experiments}

The client-style tests created two distinct synthetic or loopback peers. Trusted A and lower-trust B advertised the same target name; the test then invoked the framework's selected agent or tool and recorded which client or endpoint received the request. Controls differed by target. AutoDev changed B's remote card name; Any-Agent assigned B a distinct local tool name while both cards retained the same name; and the Google tests checked credential or callback binding with the collision intact. The preserved UiPath and BeeAI tests did not include matched distinct-name controls.

The first-party sample test did not run a model. It compiled the pinned source paths and replaced only network and SDK dependencies with inert in-process doubles, isolating name-to-route behavior from probabilistic model selection. The test was replayed on 9 September 2026. Both affected paths selected B, both one-variable controls selected A, and duplicate rejection stopped dispatch.

Solace Agent Mesh required a different test because its authority-bearing identity spans a registry, topic, and queue. The test established registry replacement and production publication to the shared name-derived route. We did not simulate a production broker topology, so final consumption by B remains conditional.

\subsection{Tests for stronger consequences}

After proving route substitution, we tested stronger consequences separately:

\begin{itemize}
  \item an A-only synthetic transport credential appearing at B's request boundary;
  \item an A-owned in-process tool becoming directly callable by B;
  \item a privileged action occurring without a new model or policy decision; and
  \item B-controlled output reaching a model that also has a host-only tool.
\end{itemize}

The fourth test confirms only that the data can reach a later decision point. Scripted models made the dataflow deterministic; their behavior does not measure exploit reliability against a production model.

\subsection{Evidence integrity and safety}

All repository revisions, test patches, source locators, and preserved results were hashed. On 9 September 2026, the SHA-256 manifests for the evidence bundle, pinned source, seven reproducer patches, and seven authority tests all passed. Runtime traffic was limited to recording doubles and study-controlled loopback peers. No third-party deployment, account, agent, or production data was contacted.

\section{Cross-Implementation Results}

\subsection{Protocol boundary}

\claim{C001}At the pinned A2A revision, \code{AgentCard.name} is required and described as a human-readable name. The card does not carry a stable agent identifier, and the discovery guidance recognizes curated registries while leaving their API and identity model outside the specification\cite{a2a-spec-98853be}. The protocol neither requires name uniqueness nor instructs a host to build \code{routes[card.name]}. This omission permits the vulnerable mapping but does not mandate it.

\subsection{Affected paths}

\claim{C003}Table~\ref{tab:targets} summarizes the seven affected paths. The tests exercise production classes with recording clients or real loopback HTTP endpoints. The Solace test instead runs the production publish path and records the broker destination before delivery. Source anchors are pinned at the exact revisions\cite{adk-python-b018062,adk-js-ac84bb4,uipath-langchain-5331f6e,beeai-0e7e3e5,solace-mesh-f71e744,any-agent-58938fe,autodev-23777fe}. The Any-Agent composition used OpenAI Agents SDK 0.20.0 under its default \code{warn} collision policy, which retains the last colliding function tool\cite{openai-agents-d2bda3f}.

\begin{table}[H]
\centering
\caption{Affected collision-to-dispatch paths at pinned revisions. ``First'' and ``last'' describe resolver precedence, not attacker privilege.}
\label{tab:targets}
\small
\begin{tabularx}{\textwidth}{@{}L{0.19\textwidth}L{0.27\textwidth}L{0.15\textwidth}Y@{}}
\toprule
\textbf{Implementation} & \textbf{Name becomes} & \textbf{Resolution} & \textbf{Observed consequence} \\
\midrule
Google ADK Python 2.8.0 & local agent name via Agent Registry helper & warning; first match & selected B agent's A2A client called \\
Google ADK TypeScript 2.0.0 & normalized local agent name via registry helper & first match & selected B agent's \code{sendMessage} called \\
UiPath LangChain 0.17.3 & generated A2A function-tool name & later replacement & surviving tool used B resource transport \\
BeeAI Framework 0.1.83 & handoff tool and requirement selector & first match & client created from B card URL and sent \\
Solace Agent Mesh 1.28.8 & card registry key, request topic, and queue identity & update/shared route & publish used collided route; consumption conditional \\
Mozilla Any-Agent 1.18.0 & \code{call\_\{card-name\}} function tool & later tool retained & JSON-RPC request reached only B loopback endpoint \\
AutoDev 2.4.1 lineage & card and client map key & later replacement & JSON-RPC request reached only B loopback endpoint \\
\bottomrule
\end{tabularx}
\end{table}

\claim{C004}In all six client-style integrations, dispatch selected B's bound client or endpoint while the legitimate A binding was not invoked. Solace instead collapsed A and B onto one broker address. The source proves publication to that shared route; queue access type, ACLs, and binding state decide whether B intercepts, becomes an active or standby consumer, or causes denial.

\subsection{Four implementation forms}

The targets reach the same invariant violation through four constructions:

\begin{enumerate}
  \item \textbf{Agent-tree ambiguity.} Google ADK retains duplicate local names and returns the first matching sub-agent.
  \item \textbf{Tool identity collapse.} UiPath, BeeAI, and Any-Agent derive tool or handoff identifiers from card names; the graph or backend then chooses one duplicate.
  \item \textbf{Client-map replacement.} AutoDev and multiple first-party samples rekey configured endpoints by card name, so a later peer replaces both card and transport.
  \item \textbf{Broker identity collapse.} Solace uses the name across discovery, registry lookup, request topics, and queue identity.
\end{enumerate}

These forms show why ``the host uses a dictionary'' is too narrow. The security property is the binding from admitted identity to final execution target, regardless of storage structure.

\subsection{Google ADK reachability boundary}

\claim{C006}The Google tests use two real remote-agent objects with duplicate local names and prove that the agent tree warns or accepts, resolves one duplicate, and invokes the selected B client. They do not, by themselves, prove remote control of the local name. In Python, direct \code{RemoteA2aAgent} construction requires a caller-supplied \code{name} documented as unique; TypeScript similarly receives the local name through \code{RemoteA2AAgentConfig}. The affected Agent Registry helpers independently derive that local name from embedded Agent Card content\cite{adk-python-b018062,adk-js-ac84bb4}.

The tests confirm unsafe duplicate resolution and wrong selected-client dispatch. Both registry integrations also contain a production card-to-name path. This study did not demonstrate that a low-privilege agent can introduce or refresh the required record in a deployed Google Cloud project. Direct-constructor misuse may be invalid operator configuration rather than a remote attack. The Google ADK result is therefore a vulnerable routing primitive with conditional attacker reachability, not an end-to-end remotely exploitable zero-day.

\section{Impact and Authority Boundary}

\subsection{What follows from wrong-peer dispatch}

\claim{C005}When the necessary conditions in Section~2 hold, B receives a request that the host selected under A's name. The request can include prompts, document fragments, task metadata, caller context, or conversation state. B can also supply a response on the selected path. That is a confidentiality and response-integrity failure whenever A and B have different authorization for the content, even if no credential crosses the boundary.
The controlled tests record the transport or endpoint that received each request. Whether B can enter that routing domain, and whether the diverted data crosses a meaningful trust boundary, remains deployment-specific.

\subsection{Credentials and delegated context}

\claim{C007}The collision usually selected B's complete client, agent, or tool closure rather than merging A's configuration into B. Across the tested ADK Python, ADK TypeScript, BeeAI, Any-Agent, and AutoDev bindings, an A-only transport credential did not appear at B. The observed UiPath bearer was shared host authentication at the managed proxy client boundary; it was not A-specific, and downstream delivery to B was not shown. The number of demonstrated direct A-specific credential transfers in these tested bindings was \DirectCredentialTransfers{}.
The shared negative result does not imply one retention mechanism. BeeAI's handoff clone rebuilt the selected agent without its original HTTP parameters, while the tested AutoDev path exposed no per-agent credential setting\cite{beeai-0e7e3e5,autodev-23777fe}.

\claim{C008}Solace differs from the client-style cases. Its publish path places the entire \code{a2aUserConfig} on the request sent to the name-derived route\cite{solace-mesh-f71e744}. That structure may contain caller identity, scopes, or delegated tokens. B receives that authority only if the deployment populated it and B can consume the collided route. Exclusive and non-exclusive queue modes change the delivery behavior\cite{solace-queue-access}. We therefore report \ConditionalDelegatedCases{} conditional delegated-context case, not a universal token leak.

\subsection{Tools and execution}

\claim{C009}No tested target transferred an A-owned in-process tool object to B or deterministically executed a privileged A or host action solely because of the name collision. UiPath and BeeAI did demonstrate \ModelMediatedChains{} second-stage dataflows: B-controlled output became model input in a context where a host-only tool was available, and a scripted model then chose that tool. This validates reachability to a new decision point. It is a compound prompt-injection path, not direct authority inheritance, and it does not measure production model reliability.

\begin{table}[ht]
\centering
\caption{Evidence boundary after the routing decision.}
\label{tab:authority}
\small
\begin{tabularx}{\textwidth}{@{}L{0.23\textwidth}L{0.20\textwidth}Y@{}}
\toprule
\textbf{Outcome} & \textbf{Status} & \textbf{Interpretation} \\
\midrule
A-intended request reaches B & Proven in six client paths & Direct confidentiality and response-integrity impact when trust differs \\
Name-derived broker route & Proven; delivery conditional & Interception, standby, or denial depends on broker controls \\
A-specific credential reaches B & Not observed & No A-specific transfer was observed; credential behavior differed by target \\
Delegated caller context & One conditional path & Solace forwards configuration, but token population and B consumption are required \\
A-owned tool or direct privileged action & Not observed & The collision alone does not grant an in-process capability \\
B output reaches a later model/tool decision & Plumbing proven in two paths & Separate model, policy, and authorization decisions determine final action \\
\bottomrule
\end{tabularx}
\end{table}

\subsection{Severity}

\claim{C015}No single severity score accurately represents every composition. If B can join a host that routes sensitive A-selected content by name, wrong-peer disclosure and response substitution can be high impact. If every peer is mutually trusted for every task, the impact may be correctness or availability. Credential and execution impact depends on attachments and sinks after routing. A report should therefore score the concrete deployment path, not the abstract existence of a duplicate name.

\section{Ecosystem Propagation and Counterexamples}

\subsection{Reference-code propagation}

\claim{C010}The first-party A2A samples contained at least \OfficialSamplePaths{} paths that fetched cards from configured addresses, stored connection objects under \code{card.name}, and later dispatched by that key\cite{a2a-samples-6603ba3}. A bounded GitHub code search for the exact assignment returned the first 100 indexed file matches across \PropagationRepositories{} distinct public repositories.
That number measures propagation, not prevalence. The set includes forks, tutorials, workshops, notebooks, and descendants of the same sample. It is neither a count of independent implementations nor a count of live deployments. It shows that the unsafe identity pattern entered reusable ecosystem reference code.

\subsection{Negative controls}

\claim{C011}We also reviewed \NegativeControls{} paths that did not exhibit the exact wrong-peer substitution behavior. Their controls were not identical, but each prevented a duplicate display name from silently choosing another transport.

\begin{table}[H]
\centering
\caption{Negative-control designs on inspected paths.}
\label{tab:negative-controls}
\small
\begin{tabularx}{\textwidth}{@{}L{0.31\textwidth}Y@{}}
\toprule
\textbf{Implementation} & \textbf{Relevant control} \\
\midrule
Microsoft Semantic Kernel .NET & duplicate function insertion fails closed\cite{semantic-kernel-872d29e} \\
Microsoft Agent Framework Python & duplicate workflow executor identifiers are rejected\cite{ms-agent-framework-540ad6b} \\
Google ADK Java & remote agents use explicit local names; duplicate sub-agent names are rejected\cite{adk-java-9bcfffd} \\
Google ADK Go & remote agents use explicit local names; the inspected multi-agent loader rejects duplicates\cite{adk-go-d56d1a8} \\
Strands Agents SDK TypeScript & exact and normalized-equivalent tool names are rejected\cite{strands-ts-00e0488} \\
\code{python-a2a} & later collisions receive distinct suffixed aliases\cite{python-a2a-40b969a} \\
Agno & routing uses an operator-controlled agent identifier rather than the card display name\cite{agno-6c5112e} \\
\bottomrule
\end{tabularx}
\end{table}

These designs are not equally strong. Suffixing prevents silent replacement but can create unstable aliases; a stable authenticated identifier is preferable. Collectively, they establish that A2A compatibility does not require wrong-peer dispatch and that implementation-level remediation is practical.

\section{Protocol, Implementation, and Deployment Responsibility}

\claim{C001}A2A supplies a required human-readable card name but no stable agent identifier or collision semantics at the pinned revision\cite{a2a-spec-98853be}. That is a protocol-level identity gap: interoperable discovery lacks a standard field and rule for the identity that should survive registration, refresh, selection, and dispatch.
\claim{C012}The name collision is not a wire-protocol attack. The protocol does not require hosts to key routes by \code{AgentCard.name}, and safe conforming implementations exist. Each affected framework owns the point where it discards an endpoint or registry distinction, accepts ambiguity, and dispatches through a fail-open resolver. Deployments separately control who may enter that resolver's domain.

\begin{table}[ht]
\centering
\caption{Responsibility follows the layer that controls each condition.}
\label{tab:responsibility}
\small
\renewcommand{\arraystretch}{1.12}
\begin{tabularx}{\textwidth}{@{}L{0.21\textwidth}L{0.35\textwidth}Y@{}}
\toprule
\textbf{Layer} & \textbf{Observed gap} & \textbf{Action} \\
\midrule
A2A protocol & no interoperable stable identity or duplicate-name semantics & declare names non-authoritative; define or profile an origin-bound stable identifier and refresh rules \\
Host/framework & display name promoted to route, tool, workflow, topic, or queue identity & preserve enrolled identity; reject ambiguity; test the final dispatch binding \\
Registry/deployer & attacker-controlled B may enter A's routing domain & authenticate publishers; scope registration and refresh; separate broker namespaces and ACLs \\
Credential/policy layer & authority may be attached after an ambiguous route resolves & resolve authenticated identity before attaching credentials, scopes, or privileged sinks \\
\bottomrule
\end{tabularx}
\end{table}

Calling the issue only a deployer misconfiguration ignores framework code that silently replaces a stable distinction with remote text. Calling it a universal protocol vulnerability ignores implementations that never perform that replacement. It is a protocol-enabled implementation vulnerability class whose reachability and impact depend on the deployment.

\section{Defenses}

\claim{C013}In the first-party sample test, the one-variable controls restored dispatch to A when B no longer collided, and duplicate rejection stopped dispatch before either peer received the request. The durable fix is to preserve identity rather than manage winner order.

Hosts, frameworks, and registries should apply the following controls:

\begin{enumerate}
  \item \textbf{Route by an enrolled identifier.} Preserve the operator- or registry-controlled ID associated with the configured endpoint. Do not rekey the object after card retrieval.
  \item \textbf{Bind identity to origin.} Associate the ID with the configured endpoint, authenticated registry subject, workload identity, or expected signing key. A stable but self-asserted ID can still be stolen.
  \item \textbf{Keep names presentational.} Names may appear in user interfaces and model context, but workflow edges, tool registries, authorization, and broker topics should resolve through the stable ID.
  \item \textbf{Fail closed on aliases.} If compatibility requires a name-only API, reject exact and normalization-equivalent duplicates. Never choose a peer by insertion or list order. Return all candidates only if a trusted caller can disambiguate by stable ID.
  \item \textbf{Attach authority after identity resolution.} Select credentials and delegated scopes from the authenticated identity, not a card-derived name. Refuse a refresh that unexpectedly changes the bound identity.
  \item \textbf{Carry identity into the broker.} Topic, queue, and consumer identity must use the stable principal. An in-memory stable registry does not help if two peers can bind the same name-derived broker route.
  \item \textbf{Test the invariant.} Register two distinct origins or resources with identical and normalization-equivalent names. Registration must fail clearly, or stable-ID dispatch must remain bound to the intended endpoint regardless of order.
\end{enumerate}

\begin{figure}[H]
  \centering
  \resizebox{\textwidth}{!}{\begin{tikzpicture}[
  node distance=8mm and 9mm,
  font=\small,
  box/.style={draw=ink,rounded corners=2pt,align=center,minimum height=9mm,inner sep=5pt,fill=white},
  stable/.style={box,draw=safe,very thick,minimum height=20mm},
  display/.style={box,draw=muted,dashed,fill=panel,minimum height=20mm},
  arr/.style={-{Latex[length=2.5mm]},thick,draw=safe},
  meta/.style={-{Latex[length=2.5mm]},draw=muted,dashed},
  note/.style={font=\scriptsize,color=muted,align=center}
]
  \node[stable,text width=31mm] (enroll) {Enrollment record\\\code{id=a-17}\\authenticated origin A};
  \node[stable,right=of enroll,text width=32mm] (binding) {Origin-bound mapping\\\code{id=a-17 -> client A}};
  \node[stable,right=of binding,text width=31mm] (route) {Workflow/API edge\\selects \code{id=a-17}};
  \node[stable,right=of route,text width=29mm] (client) {Client A\\credentials attached\\after ID resolution};

  \node[display,below=13mm of binding,text width=32mm] (name) {Agent Card display alias\\\code{payments}};
  \node[display,right=of name,text width=31mm] (ui) {UI and model context\\non-authoritative label};

  \draw[arr] (enroll) -- (binding);
  \draw[arr] (binding) -- (route);
  \draw[arr] (route) -- (client);
  % Presentation metadata branches away from the authority path. The final
  % segment enters the alias box horizontally, keeping the routing semantics
  % visually distinct from the solid identity path above.
  \draw[meta] (enroll.south) |- (name.west);
  \draw[meta] (name) -- (ui);

  \coordinate (displaybottom) at ($(name.south)!0.5!(ui.south)$);
  \node[note,below=7mm of displaybottom,text width=75mm] {Duplicate aliases are rejected or made explicitly ambiguous; they never replace the stable route.};
\end{tikzpicture}}
  \caption{Safe routing keeps the enrollment identity on the authority path and moves display names outside it.}
  \label{fig:safe-binding}
\end{figure}
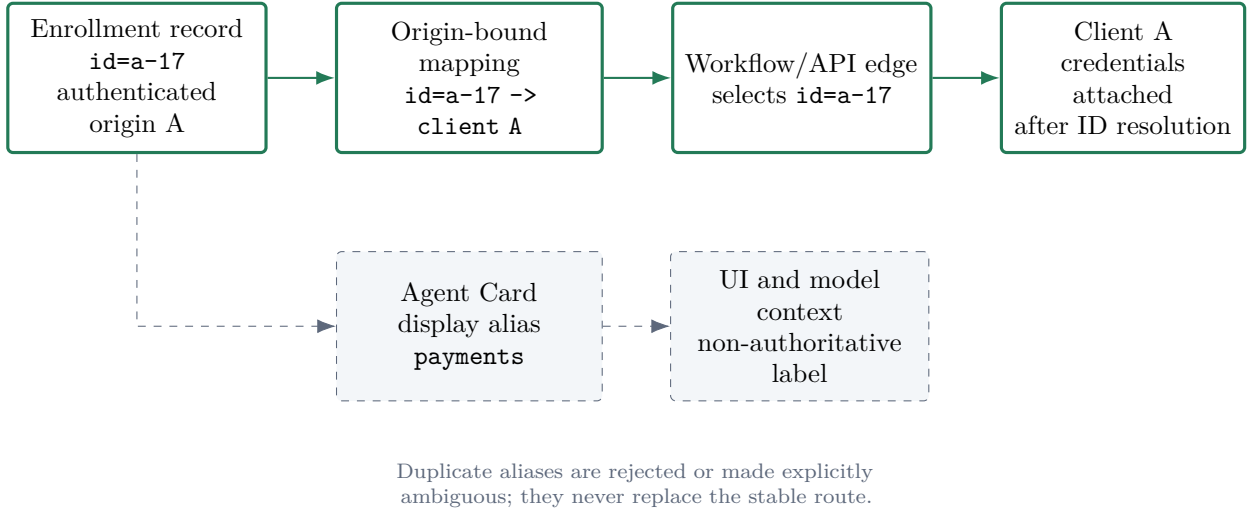

A warning is not a security control if the ambiguous graph remains usable. Similarly, reversing insertion order only changes the winner. Remediation is complete when an untrusted card field can no longer select another principal's transport or authority.

\section{Limitations and Ethical Scope}

\claim{C005}The attack model begins after B is admitted or compromised. We did not establish a universal enrollment bypass. Actual exploitability depends on registry permissions, discovery topology, configuration ownership, and broker ACLs, as well as the name collision itself.

\claim{C006}Google ADK requires special caution. The focused tests manually supplied duplicate local names, while the Agent Registry helpers provide the source-established card-to-name path. This study did not test a Google Cloud project or prove that a low-privilege agent can create or refresh the needed registry record. That missing admission edge prevents an end-to-end remote-exploit claim for Google ADK.

Solace final delivery was not exercised in a production broker topology. The source and focused test establish registry replacement, route derivation, and publication of caller configuration to the collided route. Exclusive versus non-exclusive queues, bind order, and ACLs determine whether B receives the message.

\claim{C015}The authority study used synthetic credentials and harmless action counters. It did not demonstrate production secrets, business actions, or one universal severity. Scripted-model tests prove that B's output can reach another decision point; they do not estimate whether a real model, prompt, approval rule, or guardrail would invoke a privileged tool.

\claim{C010}The seven-target corpus was purposive. Package popularity and repository search were used to find and contextualize implementations, not to estimate a population rate. The \PropagationRepositories{}-repository exact-pattern snapshot cannot be converted into vulnerable products, exposed installations, or independent codebases.

\claim{C014}The prior-art review covered public, indexed material available by 9 September 2026. Private reports, renamed issues, and unindexed content may exist. We found no earlier public description of the complete mechanism within those bounds; this is not a claim of novelty or priority.

All runtime traffic was confined to in-process doubles or study-controlled loopback peers. No third-party endpoint or account was probed. Results report exact pinned revisions rather than present-tense vulnerability status.

\section{Directly Related Work}

\claim{C014}The closest protocol work is A2A issue 1014, which states that \code{agent\_card.name} is not guaranteed unique and is vulnerable to branding changes, and proposes a persistent identifier\cite{a2a-issue-1014}. That work identifies the missing primitive. It does not document the complete path from duplicate display name through a fail-open host resolver to deterministic wrong-peer dispatch.

LevelBlue's Agent-in-the-Middle experiment shows that a compromised peer can exaggerate Agent Card capabilities so a model selects it for tasks\cite{neaves-agent-in-middle}. The attacker capability and confidentiality consequence are adjacent, but the mechanism is model-mediated description steering. The collision studied here instead corrupts a deterministic local identity binding after peers are admitted and can be tested without model choice.

Palo Alto Networks' A2A security guidance discusses agent impersonation and Agent Card shadowing, including similar names, capabilities, identities, and endpoints\cite{panw-a2a-shadowing}. That establishes broad shadowing and impersonation as prior art. The contribution here is narrower: cross-implementation evidence for exact or normalized name collision entering an authority-bearing resolver and reaching the wrong transport, tool, or broker route.

CWE-706 provides the closest general software-weakness category: a name or reference resolves to a resource outside the intended control sphere\cite{cwe-706}. Agent-name collision applies that pattern to multi-agent discovery, where a remote presentation field displaces an enrolled principal at dispatch.

These works rule out broad novelty claims about malicious Agent Cards, agent impersonation, name collisions, or task capture. The bounded search found no earlier public description of the complete chain from duplicate name, through fail-open resolution, to wrong-peer dispatch.

\section{Conclusion}

\claim{C002}Agent-name collision is a security vulnerability when a host discards a stable admission distinction, promotes remote presentation metadata into routing identity, and silently resolves an ambiguity to another principal. Seven independently maintained implementation paths reached that failure through agent lookup, tool registration, client maps, and broker naming.

\claim{C012}Responsibility spans three layers. The protocol leaves an identity-semantics gap, affected frameworks implement the unsafe resolver, and deployments determine who can enter the routing domain and what authority follows the selected route. Framework behavior makes this more than an intentional operator misconfiguration, while safe conforming implementations rule out a universal A2A protocol exploit.

The proven common impact is wrong-peer dispatch. Stronger outcomes require separate evidence. The tested client bindings did not transfer A-specific credentials or A-owned tools, while one broker path conditionally forwarded delegated caller context and two framework paths exposed a separate model-mediated chain.
\claim{C013}Remote presentation metadata must not select the security principal or transport binding. Preserving an origin-bound stable identity, rejecting ambiguous aliases, and attaching credentials only after identity resolution removes the measured failure without constraining human-readable names.

\appendix
\section{Reproducibility Record}

\claim{C003}The study pinned one revision per tested path and retained the patches, exact commands, result transcripts, and SHA-256 manifests. Table~\ref{tab:repro} records the principal executions; Table~\ref{tab:targets} and the bibliography provide the source locations.
The public artifact repository contains the manuscript source, normalized results, pinned source maps, focused reproducer patches, recorded outputs, and an archive-integrity verifier\cite{agent-name-collision-artifacts}.

\begin{longtable}{@{}L{0.25\textwidth}L{0.25\textwidth}L{0.44\textwidth}@{}}
\caption{Focused test record.}\label{tab:repro}\\
\toprule
\textbf{Target} & \textbf{Evidence type} & \textbf{Recorded result} \\
\midrule
\endfirsthead
\toprule
\textbf{Target} & \textbf{Evidence type} & \textbf{Recorded result} \\
\midrule
\endhead
Google ADK Python & production agents and recording clients & two tests passed; duplicate warning retained \\
Google ADK TypeScript & production agent tree and A2A clients & two tests passed \\
UiPath LangChain & production tool construction and graph node & three tests passed; one warning \\
BeeAI Framework & A2A agent, handoff tool, and scripted model & one focused test passed \\
Solace Agent Mesh & production registry and publish method & two tests passed; broker delivery not asserted \\
Mozilla Any-Agent & two loopback A2A peers; OpenAI Agents SDK 0.20.0 (\code{warn}) & collision and local tool-name override tests passed; two warnings \\
AutoDev & production Kotlin consumer with loopback HTTP peers & collision and distinct-name controls passed \\
First-party samples & pinned source paths with inert doubles & two affected paths, two distinct-name controls, and duplicate rejection passed \\
\bottomrule
\end{longtable}

\claim{C004}All six client-style collision tests recorded selection of B's client or endpoint. Controls differed by target: AutoDev changed B's remote card name; Any-Agent used a local tool-name override; and the Google tests checked credential or callback binding with the collision intact. The preserved UiPath and BeeAI tests did not include matched distinct-name controls. The separate first-party sample test changed only B's name and exercised duplicate rejection. The broker test stops at the production publish boundary, so final peer consumption remains conditional.

\subsection{Evidence integrity}

The archived results, source corpus, collision-test patches, and authority-boundary test patches each had an independent manifest. All four manifests passed during verification on 9 September 2026. The deterministic sample test also rechecked its pinned source hashes before execution. The reported counts are generated from a structured results file rather than re-entered manually.

\subsection{Reproduction safety}

Reproduction requires only disposable local checkouts, recording clients, or loopback HTTP peers. Tests should use synthetic messages and credentials. No external agent, shared broker, production registry, or real user data is required to confirm the resolver behavior. Testing a deployment's admission or broker-consumption edge requires explicit authorization and is outside this artifact.

\section*{Generative AI Usage}

To help manage and cross-check the repository-scale evidence corpus, OpenAI Codex
assisted with repository inspection, orchestration and checking of experiment
scripts, evidence normalization, citation-metadata collection, drafting and
revision of manuscript text and \LaTeX{} figures, and build and visual verification.
The author determined the research questions, scope, methods, evidentiary
boundaries, claims, and conclusions; reviewed the cited sources, experimental
records, controls, citations, and final manuscript; and made all substantive
decisions. Codex outputs were not treated as independent evidence or validation.
The author takes full responsibility for the paper.

\bibliographystyle{plain}
\begingroup
\small
\raggedright
\bibliography{references}
\endgroup

\end{document}